\documentclass[%
 reprint,
 superscriptaddress,
 amsmath,amssymb,
 aip,
 jcp,
floatfix,
]{revtex4-1}

\usepackage{graphicx}
\usepackage{amsmath}
\usepackage{amssymb}

\usepackage{graphicx}% Include figure files
\usepackage{dcolumn}% Align table columns on decimal point
\usepackage{bm}
\usepackage{tabularx}
\usepackage{url} % or \usepackage{hyperref}

\draft % marks overfull lines with a black rule on the right
\DeclareUnicodeCharacter{2064}{}

\usepackage{bbm}
\usepackage{color}
\usepackage[normalem]{ulem}

\newcommand{\todo}[1]{{\color{red}{{\textbf{\large #1}}}}}

\begin{document}

\title{
Ions at electrochemical interfaces: from explicit to implicit molecular solvent descriptions} %Title of paper

\author{Swetha Nair}
\affiliation{Sorbonne Universit\'e, CNRS, Physicochimie des \'Electrolytes et Nanosyst\`emes Interfaciaux, F-75005 Paris, France}

\author{Guillaume Jeanmairet}
\affiliation{Sorbonne Universit\'e, CNRS, Physicochimie des \'Electrolytes et Nanosyst\`emes Interfaciaux, F-75005 Paris, France}
\affiliation{R\'eseau sur le Stockage Electrochimique de l'Energie (RS2E), FR CNRS 3459, 80039 Amiens Cedex, France}

\author{Benjamin Rotenberg}
\affiliation{Sorbonne Universit\'e, CNRS, Physicochimie des \'Electrolytes et Nanosyst\`emes Interfaciaux, F-75005 Paris, France}
\affiliation{R\'eseau sur le Stockage Electrochimique de l'Energie (RS2E), FR CNRS 3459, 80039 Amiens Cedex, France}
\email{benjamin.rotenberg@sorbonne-universite.fr}

\date{\today}

\begin{abstract}
We investigate the interplay between electronic screening inside a metal and screening by a polar molecular solvent, focusing on their impact on the charge induced by an ion and the solvent structure at the interface. To that end, we consider atomistically resolved electrodes within the Thomas-Fermi model of screening and describe the molecular solvent either explicitly via classical molecular dynamics or implicitly using Molecular Density Functional Theory (MDFT). Specifically, we examine the effect of screening by tuning the Thomas-Fermi screening length $l_\text{TF}$, the ion charge by considering Na$^+$ and Cl$^-$ and the solvent nature by studying water and acetonitrile. Consistently with our previous findings without solvent, $l_\text{TF}$ significantly affects the charge distribution inside the metal. However, $l_\text{TF}$ has no significant impact on the interfacial solvent structure, suggesting that its effect on the charge distribution induced inside the metal by the ion is essentially due to how the metal responds to the (same) external charge distribution, including the solvent, even though the coupling between both sides of the interface may play a secondary role. Furthermore, MDFT accurately reproduces fine details of the interfacial solvent structure around the ion at a fraction of the computational cost of MD simulations. These results highlight the relevance of MDFT as a powerful tool to model electrochemical systems at the molecular level.
\end{abstract}


\maketitle %\maketitle must follow title, authors, abstract and \pacs


%%%%%%%%%%%%%%%%%%%%%%%   %%%%%%%%%%%%%%%%%%%%%%%
\section{Introduction}

The behavior of electrode-electrolyte interfaces is crucial for various electrochemical applications including energy storage and electrocatalysis. Their properties arise from the interplay between the metal's electronic response to charges from ions and solvent molecules, and the subsequent reorganization of these species based on the charge distribution at the metal surface. The Thomas-Fermi model~\cite{thomas1927a,fermi1927a} provides a simple description of electronic screening within the metal over a characteristic length, $l_\text{TF}$, determined by the electronic density of states at the Fermi level~\cite{AshcroftMermin}. On the electrolyte side, the screening of electrostatic interactions between ions in polar solvents is characterized by the Bjerrum length, while ionic screening in dilute electrolytes is governed by the Debye screening length. The interplay between these screening mechanisms plays a significant role in shaping the interfacial properties, including induced charge, ion-ion interactions, electric double layer structure, capacitance, and the overall thermodynamics and kinetics of electrochemical processes.

The mutual influence of metal, solvent and ions have been extensively studied within continuum electrostatics, notably though not only by Kornyshev and co-workers~\cite{kornyshev_image_1977,kornyshev1978a,vorotyntsev_1980,kornyshev1982a,dzhavakhidze_activation_1987,phelps1990a,luque2012a,kornyshev2013a,rochester2013a,comtet2017a,Kaiser_2017}. An important finding was the establishment of a characteristic length for the lateral decay of the charge/potential induced in the metal by an ion scaling as $l_\text{TF}\epsilon_\textrm{solv}/\epsilon_\textrm{met}$, \textit{i.e.} an increase of the screening length by the ratio between the solvent permittivity and that of the Thomas-Fermi metal (since usually $\epsilon_\textrm{solv}>\epsilon_\textrm{met}$). In parallel, the charge distribution of the metal in response to an external charge was studied using electronic Density Functional Theory (DFT) with both simplified (1D) geometries~\cite{Lang-jellium-1973,schmickler_interphase_1984,Smith-jellium-1989} and with more advanced atomically resolved surfaces~\cite{jung_self-consistent_2007,yu_abinitio_2008}. Recent advances in simulation techniques have further improved the modeling of electrolyte interfaces, incorporating polarization effects between media of different permittivities and explicitly accounting for ions~\cite{tyagi_icmmm2d_2007,tyagi_electrostatic_2008,arnold2013a,girotto_simulations_2017,nguyen_incorporating_2019,maxian_fast_2021,yuan_particleparticle_2021,dos_santos_modulation_2023}. Both \emph{ab initio}~\cite{sakong_electric_2018,elliott_qmmd_2020,le_modeling_2021,sakong_structure_2022,takahashi_accelerated_2022,sundararaman_improving_2022,grisafi_predicting_2023} and classical molecular dynamics~\cite{Siepmann_1995,spohr_molecular_1999,kornyshev_electrochemical_2007,Reed_2008,fedorov2014a,onofrio_voltage_2015,breitsprecher_electrode_2015,liang_applied_2018,nakano_chemical_2019,takahashi_unified_2022,sitlapersad_simple_2024,scalfi_molecular_2021,jeanmairet_microscopic_2022} have provided key insights into electrode-electrolyte interfaces and the impact of electrolyte-induced charge within the metal on interfacial properties~\cite{willard2009a,limmer2013b,willard2013a,paek2015a,buraschi2020a,serva_effect_2021,loche_effects_2022,ntim_role_2020,ntim_molecular_2023,ntim_differential_2024}. 

Recent approaches have also extended the description of Thomas-Fermi metals to account for electronic screening and its effects on interfacial thermodynamics in classical molecular simulations~\cite{Scalfi_tfmodel_2020,schlaich_electronic_2021,scalfi-pnas-2021}. The metallicity of an electrode, which indicates its electronic structure, also influences electron transfer kinetics, as illustrated by recent experiments demonstrating that the kinetics of heterogeneous electron transfer reactions at well-defined twisted bilayer graphene (TBG) surfaces can be adjusted by varying the interlayer Moiré twist angle between the two layers~\cite{yu_tunable_2022}. Furthermore, recent molecular simulations combined with continuum dielectric theory have provided deeper insights into these observations, establishing a link between the twist angle and the Thomas-Fermi screening length through the density of states (DOS) of TBG~\cite{coello2024microscopic}. This relationship between DOS and Thomas-Fermi screening length has also been investigated in classical molecular simulations~\cite{goloviznina2024accounting}, specifically tuning the description of the charge density around electrode atoms in fluctuating charge models. Recently, we investigated the charge induced in a perfect metal ($l_\text{TF}$ = 0) by an ion near an atomically resolved gold electrode~\cite{pireddu_molecular_2021} and extended this study to include the screening effect within the metal, examining the charge induced by an ion in vacuum in a Thomas-Fermi metal using a constant potential fluctuating charge model~\cite{Nair-2024}. We described the influence of the Thomas-Fermi screening length ($l_\text{TF}$) and the ion's position relative to a graphite electrode on the induced charge distribution, comparing atomistic results with predictions from continuum electrostatics. 

In the present work, we further explore the interplay between electronic screening inside the metal and screening by a polar molecular solvent and their consequences on the charge induced by an ion and the solvent structure at the interface. To that end, we employ the Thomas-Fermi model for an atomistically resolved electrode, and describe the molecular solvent either explicitly using classical molecular dynamics (MD) or implicitly within classical Density Functional Theory (DFT). While classical MD simulations provide a more detailed description, sampling the equilibrium properties entails a high computational cost. Classical DFT, whereby the solvent is described implicitly via its local density instead of explicit molecules, offer a less demanding alternative, but rely on approximations typically built on Fundamental Measure Theory of hard-spheres\cite{rosenfeld_theory_1979} that prevent a realistic description of molecular solvents~\cite{jeanmairet_microscopic_2022}. Molecular Density Functional Theory (MDFT) provides an alternative route~\cite{Ramirez-2002,zhao-2011,Borgis-2012,Jeanmairet-MDFTwater-2013,Jeanmairet-2013,Jeanmairet-2015,Jeanmairet-2016,Ding-2017,Jeanmairet-2019,Jeanmairet-rsc-2019,Borgis-2021,jeanmairet_microscopic_2022,Hsu-Marcus-2022,Borgis-2023}, based on a Taylor expansion of the functional, which allows to describe molecular systems. Building on our previous study of a model a water-graphene capacitor with polarizable electrode without considering the role of the Thomas-Fermi screening length~\cite{Jeanmairet-2019}, here we compare MDFT with MD to study the charge induced by an ion in a solvent within a Thomas-Fermi metal. To that end, we consider the Na$^+$ cation and Cl$^-$ anion in water and acetonitrile, near a graphite electrode with varying Thomas-Fermi screening length. The various methods and considered systems are presented in Section~\ref{sec:methods}, while the results pertaining to the effects of the screening length on the induced charge distribution and on the solvent structure are provided in Sections~\ref{sec:results:inducedq} and~\ref{sec:results:solvstr}, respectively.

%%%%%%%%%%%%%%%%%%%%%%%   %%%%%%%%%%%%%%%%%%%%%%%
\section{Methods and System}
\label{sec:methods}

We first describe the Thomas-Fermi model used in constant-potential classical MD simulations in Section~\ref{sec:methods:MD} and molecular DFT in Section~\ref{sec:methods:MDFT}. We then present the system considered in Section~\ref{sec:methods:details}, before introducing the observables of interest in Section~\ref{sec:methods:observables}.

%%%%%%%%%%%%%%%%%%%%%%%  
\subsection{Thomas Fermi model within constant-potential classical MD simulations}
\label{sec:methods:MD}

In recent years, fluctuating-charge models have been widely used for simulating electrochemical systems within classical MD simulations involving electrodes maintained at a fixed potential. First proposed by Siepmann and Sprik to study water/electrode interfaces with classical MD~\cite{Siepmann_1995}, these models were adapted to electrochemical cells~\cite{reed2007a} and more recently extended to capture Thomas-Fermi screening~\cite{Scalfi_tfmodel_2020,scalfi-pnas-2021}. Within this framework, each electrode atom $i$, positioned at $\bm{r}_i$ and bearing charge $q_i$, contributes to the total charge density as:
\begin{equation} \label{eq:rho3D}
    \rho_\text{elec}({\bm r}) = \sum\limits_{i\in \text{elec}}\frac{q_i}{(2\pi w^2)^{3/2}}e^{-|{\bm r}-{\bm r}_i|^2/2w^2} \; ,
\end{equation} 
where $w$ is a fixed width of the Gaussian distribution around each atom. The instantaneous charges of all electrode atoms depend on the external charge distribution, \textit{i.e.} the positions $\bm{r}^N$ of atoms in the electrolyte. Specifically, for each configuration of the electrolyte they are determined by the constraints of constant potential imposed on each atom and of global electroneutrality of the system~\cite{scalfi_charge_2020,scalfi_molecular_2021}. The computation of the set of electrode charges $\bm{q}=\{q_i\}_{i\in \text{elec}}$ involves the potential energy of the system, which owing to the form of electrostatic interactions can be written as:
\begin{equation} \label{eq:utf}
    U(\bm{r}^N,\bm{q})= \frac{\bm{q}^{T} {\bf A}\bm{q}}{2} - \bm{q}^{T} {\bf{B}}(\bm{r}^N) + C(\bm{r}^N)
    \; ,
\end{equation}
where the first term accounts for electrostatic interactions between electrode atoms, the second for those between electrode and electrolyte, and the third for all other interactions. The function $C$ is independent of electrode charges $\bm{q}$ and includes electrostatic interactions within the electrolyte, as well as all non-electrostatic (\emph{e.g.} Lennard-Jones) interactions of the classical force field. Both the matrix ${\bf{A}}$ and the vector $\bf{B}$ account for the Gaussian nature of the charge distribution of electrode atoms and the periodic boundary conditions (PBC) employed in molecular simulations of condensed matter systems. While this description was initially introduced for perfect metals, Scalfi \emph{et al.} extended it to the Thomas-Fermi model, by including the kinetic energy of the free electron gas as in the original Thomas-Fermi approach. By making further approximations to the description of the local density in the fluctuating charge approach, the Thomas-Fermi model was recast in the same form (Eq.~\ref{eq:utf}) simply by introducing an additional term in the matrix describing electrode-electrode interactions, namely~\cite{Scalfi_tfmodel_2020}:
\begin{equation}\label{eq:ltf}
    {\bf{A}}(l_\text{TF}) = {\bf{A}}_{0} + \frac{l_\text{TF}^{2} \bar{\rho}_{at}}{\epsilon_{0}} {\bf{I}}
    \; ,
\end{equation}
where $\bf{A}_{0}$ is the matrix for $l_\text{TF} = 0$, $\bar{\rho}_{at}$ the atomic density of the electrode, $\epsilon_0$ the vacuum permittivity and $\bf {I}$ is identity matrix. This description, already implemented in several simulation packages~\cite{marin-lafleche_metalwalls_2020,coretti_metalwalls_2022,ahrens-iwers_electrode_2022} allows to tune the metallic character of the metal without additional computational cost.

%%%%%%%%%%%%%%%%%%%%%%%  
\subsection{Molecular Density Functional Theory (MDFT)}
\label{sec:methods:MDFT}

One of the objectives of the present work is to assess the quality of Molecular Density Functional Theory (MDFT) to model the solvent in which the electrodes are immersed. Two particular properties should be accurately reproduced: the solvent structure and the induced charges. MDFT is based on the ansatz of classical DFT, a theory that is most naturally formulated in the grand-canonical ensemble. It states that for a given external potential $V$, there exists a unique functional $\Omega_V$ of the classical particle density $\rho$. At its minimum, which is reached for the equilibrium particle density $\rho_\text{eq}$, this functional is equal to the grand potential~\cite{Mermin-PR-65,Evans-AP-79}, where the subscript highlights the dependence on the external potential. In MDFT~\cite{Jeanmairet-MDFTwater-2013,Ding-2017,Jeanmairet-rsc-2019,jeanmairet_microscopic_2022}, which was originally designed to study solvation problems, the working functional is actually the difference between the grand potential functional and the grand potential of the homogeneous system $\Omega_0$, which is obtained in the absence of an external perturbation  
\begin{equation} \label{eq:F=Omega-Omega0}
    F[\rho]=\Omega_V[\rho]-\Omega_0
\end{equation}
In the present work, the classical particle density represents the solvent \textit{i.e} either water or acetonitrile. Since we consider rigid molecular models (see Section~\ref{sec:methods:details}), knowledge of the position of the center of mass $\bm{r}$ and the absolute orientation $\bm{\Omega}=(\theta,\phi,\psi)$ of a molecule is sufficient to fully characterize the atomic coordinates. Consequently, the molecular solvent density $\rho(\bm{r},\bm{\Omega})$ also depends on the position and orientation. The external potential, which causes the solvent density to depart from homogeneity, is generated by the electrodes and by, if present, a cation or an anion. The functional of Eq.~\ref{eq:F=Omega-Omega0} is usually decomposed into the sum of several terms:
\begin{equation} \label{eq_F=Fid+...}
    F[\rho(\bm{r},\bm{\Omega})]=F_\text{id}[\rho(\bm{r},\bm{\Omega})]+F_\text{exc}[\rho(\bm{r},\bm{\Omega})]+F_\text{ext}[\rho(\bm{r},\bm{\Omega})].
\end{equation}

The ideal term $F_\text{id}$ is the functional associated with the corresponding non-interacting fluid, while the excess term $F_\text{exc}$ originates from the interaction between the solvent molecules. As opposed to the ideal term, which is exactly known, the excess term requires some approximation. In this work, the excess functional is expressed as the sum of the so-called hypernetted chain (HNC) functional and a bridge functional. The HNC functional, corresponds to a Taylor expansion of the excess functional around a homogeneous reference solvent density, truncated at seconder order. Even at this quadratic order, the computation of the HNC functional is not a trivial task. Indeed, it requires the knowledge of the two body correlation function of the homogeneous solvent $c^{(2)}(r,\bm{\Omega},\bm{\Omega}^\prime)$ that depends on a distance and two orientations. Moreover, its calculation is computationally challenging since it involves a spatial and an orientational convolution. The expression of the HNC functional and details about its practical implementation can be found in our previous work~\cite{Ding-2017}. The bridge functional contains in principle all terms beyond the quadratic order and cannot be computed in practice. It is approximated here using a simple, angular independent, weighted density approximation (WDA) functional~\cite{Borgis-2021}. 

The last term of Eq.~\ref{eq_F=Fid+...} is the external functional, which represents here the interaction between the solvent particles and the external potential $V$ generated by the electrodes and, if present, the cation or anion. It reads 
\begin{equation}\label{eq:Fext_q}
    F_\text{ext}[\rho(\bm{r},\bm{\Omega})]=\iint \rho(\bm{r},\bm{\Omega})V(\bm{r},\bm{\Omega}){\rm d}\bm{r}{\rm d}\bm{\Omega}.
\end{equation}
In the present case, $V$ is decomposed as the sum of a Lennard-Jones term $V_\text{LJ}$, an electrostatic term due to fixed charges $V_\text{ES,fixed}$ and an electrostatic term due to the fluctuating electrode charges $V_\text{ES,fluct}$
\begin{equation}\label{eq:Fext_q2}
   V(\bm{r},\bm{\Omega})=V_\text{LJ}(\bm{r},\bm{\Omega})+V_\text{ES,fixed}(\bm{r},\bm{\Omega})+V_\text{ES,fluct}(\bm{r},\bm{\Omega};\bm{q}).
\end{equation}
The external potentials due to the Lennard-Jones and fixed charges (cation or anion) are computed only once, at the beginning of the computation. For a given set of electrode charges, $\bm{q}$, it is possible to compute $V_\text{ES,fluct}$. Minimizing  the functional of equation \ref{eq_F=Fid+...}, we obtain the equilibrium solvent density $\rho_\text{eq}(\bm{r},\bm{\Omega})$. Since the solvent density is inhomogeneous, the corresponding charge density, $\rho_\text{c}$, does not vanish
\begin{equation}\label{eq:rho_c}
   \rho_\text{c}(\bm{r},\bm{\Omega})=\iint \rho_\text{eq}(\bm{r}^\prime,\bm{\Omega}) \sigma(\bm{r}-\bm{r}^\prime,\bm{\Omega}){\rm d}\bm{r}^\prime {\rm d}\bm{\Omega}.
\end{equation}
In Eq~\ref{eq:rho_c}, $\sigma(\bm{r},\bm{\Omega} )$ is the charge density of a water molecule with orientation $\bm{\Omega} $ located at the origin, $\bm{r}=0$. 
The solvent charge density, in turn, impacts the electrostatic potential felt by the electrode atoms. In practice, the elecrostatic potential generated by the solvent charges ($\rho_\text{c}$ discretized on a grid) enters the vector $\bf{B}$ in Eq.~\ref{eq:utf}. Following the procedure described in previous work, we perform successive functional minimization and electrode charge computation. At convergence, we obtain the equilibrium solvent density and the electrode charges fulfilling both the electroneutrality and fixed potential conditions.

%%%%%%%%%%%%%%%%%%%%%%%   
\subsection{Computational details}
\label{sec:methods:details}

\begin{figure}[!ht]
\center
\includegraphics[width=3.5in]{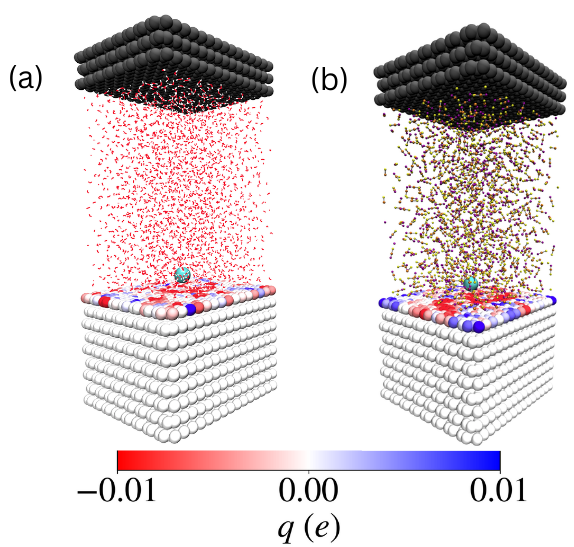}
\caption{
Snapshots of typical molecular configurations considered in his work, with water (a) or acetonitrile (b) as solvent. Both panels include a graphite electrode with explicit atoms, an ion (here a Na$^+$ cation positioned at $z_{ion} =5.1$~\AA\ from the first atomic plane of the electrode, shown in cyan), and an insulating wall with the same graphite structure (shown in dark grey). The color of the electrode atoms represents the charge induced by the ion and solvent molecules, with red and blue indicating negative and positive values, respectively. For a Thomas-Fermi screening length of $l_\text{TF} = 0$~\AA, as shown here, almost no charge is found beyond the first atomic plane in contact with the liquid. Periodic boundary conditions are applied in the $x$ and $y$ directions.
}
\label{fig:snapshot}
\end{figure}

In order to study the charge induced in a Thomas-Fermi metal by an ion in a molecular solvent, we have considered the two systems depicted in Fig.~\ref{fig:snapshot}. Both contain a metallic graphite electrode, held at a fixed potential and a insulating graphite electrode. The metallic electrode consists of 9 graphite planes with a total of 4320 atoms (to allow for the possibility of spreading the induced charged within the electrode when the screening length is larger than the interplane distance), while the insulating wall is made of 3 atomic layers only, with a total of 1440 carbon atoms. The interatomic distance within each plane is $d_{CC} = 1.42$~\AA, and the spacing between planes is $a = 3.354$~\AA. The width of the Gaussian atomic charge entering Eq.~\ref{eq:rho_c} is  $w= 0.4$~\AA, and the Thomas-Fermi length $l_\text{TF}$ ranges from $0$~\AA\, to $5$~\AA. Both systems contain a Na$^+$ or Cl$^-$ ion either in water (2160 molecules) or in acetonitrile (789 molecules). 

Periodic boundary conditions are used in the directions parallel to the walls, with lateral box dimensions of $L_x = 34.10$~\AA, $L_y = 36.92$~\AA\. The system is not periodic in the $z$ direction. The distance between the electrode and insulating wall planes in contact with the pure liquid was equilibrated for 0.5~ns in the $NPT$ ensemble at 1~atm by applying a constant force to the electrodes (pistons) with $l_\text{TF} = 0~$\AA, yielding distances of $L_z=$55.088 and $L_z=$54.070~\AA\ for the water and acetonitrile systems, respectively. The ion is fixed at a distance  $z_\text{ion}=5.1$~\AA~from the  electrode plane, above a C atom. This distance was chosen based on equilibrium density profiles of Na$^+$ ion near the graphite electrode surface in a 1~M NaCl aqueous solution (see Fig.A1 in the appendix of our previous work~\cite{Nair-2024}). In order to investigate the effect of the ion charge and of the solvent, we used the same positions for the Cl$^-$ anion, both in water and in acetonitrile. Finally, we also performed simulations of the same systems in the absence of ion.

Electrostatic interactions are computed using 2D Ewald summation method taking into account the Gaussian distribution of electrode atoms and the ion’s point charge~\cite{coretti_metalwalls_2022}. Non-electrostatic interactions are introduced through truncated and shifted Lennard-Jones (LJ) potentials, with parameters from Ref.~\citenum{Dang-jacs-1995} for Na$^+$ and Cl$^-$ and from Ref.~\citenum{Werder-jpcb-2003} for carbon, with Lorentz-Berthelot mixing rules. Water molecules are modeled using the rigid SPC/E force field~\cite{Berendsen-SPC-1987}, while acetonitrile (MeCN) molecules are described by a rigid coarse-grained three-site model involving a unified description of the methyl group~\cite{Edward-1984, Merlet-2013-acn}. The partial charges of the SPC/E water model are -0.8476$e$ for oxygen and +0.4238$e$ for hydrogen, with $e$ the elementary charge; for the three-site MeCN model, they are -0.398$e$ for nitrogen, +0.129$e$ for the central carbon and +0.269$e$ for the methyl group. Simulations are performed in the $NVT$ ensemble for 8~ns with a time step of 2~fs, using a Nos\'e-Hoover chain thermostat with a time constant of 1~ps to maintain a temperature of 298~K. All simulations were performed with the molecular dynamics code Metalwalls~\cite{marin-lafleche_metalwalls_2020}, and electrode charges were calculated at each time step via matrix inversion~\cite{scalfi_charge_2020}, taking into account the constraint of overall electroneutrality (which implies that the total charge of the electrode compensates that of the ion) in addition to the constant-potential one. Electrode charges and solvent configurations were sampled every 100 steps (0.2~ps) and 1000 steps (2~ps), respectively, for subsequent analysis.

The same systems were studied using MDFT by removing the explicit solvent molecules and employing the functional introduced in Eq~\ref{eq_F=Fid+...} to implicitly account for the solvent. The functional was minimized within a $34.10\times36.92\times120~$\AA$^3$~simulation cell, discretized on a  $102*111*120$ grid, corresponding to a resolution of approximately $0.33$~\AA~in each direction. To address the 3D PBC used in MDFT, a hard wall perpendicular to the electrode surface was introduced to avoid self interaction. This wall also ensures that the interlayer space between graphite sheets remained free of water molecules. The three Euler angles $(\theta,\phi,\psi)$ were discretized into 196 orientations,  corresponding to a truncation of the expansion of the HNC functional to $m_\text{max}=3$ (see Ref.~\citenum{Ding-2017}). For the aqueous system, an angle-independent Weighted Density Approximation (WDA) bridge functional was added to the HNC functional~\cite{borgis_simple_2020,Borgis-2021}. However, as the WDA bridge functional has not been parameterized for acetonitrile, the excess functional in this case was limited to the HNC functional.

%%%%%%%%%%%%%%%%%%%%%%% 
\subsection{Observable properties}
\label{sec:methods:observables}

In order to analyze the charge distribution inside the metal and the structure of the solvent at the interface, in the following sections we will consider several observables. The equilibrium charge density is obtained by averaging Eq.~\ref{eq:rho3D} over the configurations in the canonical ensemble, which simply amounts to replacing the instantaneous charges $q_i$ (which depend on the microscopic configuration of the liquid) by the ensemble average $\left\langle q_i \right\rangle$. It is more conveniently analyzed by integrating over the depth of the electrode to obtain the 2D charge density 
\begin{equation} \label{eq:2d}
    \sigma_\text{ind}(x,y) = \int_{-\infty}^0 \rho_\text{elec}(x,y,z) \,{\rm d}z
    \; .
\end{equation}
In practice, it is more efficient to perform the integration over $z$ analytically and directly reconstruct the 2D distribution as
\begin{equation} \label{eq:2d:MD}
    \sigma_\text{ind}(x,y) = \sum\limits_{i\in \text{elec}}\frac{\left\langle q_i \right\rangle}{2\pi w^2}e^\frac{-[(x-x_i)^2+(y-y_i)^2]}{2w^2} \; .
\end{equation} 
This quantity is computed on a 2D grid with a bin width of $0.1$~\AA,\, where the charge in each bin is calculated using the analytical integral of the 2D Gaussian distributions (see Ref.~\citenum{Nair-2024}). In order to analyze the radial decay away from the ion located at $(x,y,z)=(0,0,z_\text{ion})$, we also introduce the radial charge density:
\begin{equation} \label{eq:rad}
    \sigma_{ind}(r) = \frac{1}{2\pi r}\iint \sigma_{ind}(x,y)\delta\left(r-\sqrt{x^2 + y^2}\right) \,{\rm d}x{\rm d}y
    \; ,
\end{equation}
with $\delta$ Dirac's delta function as well as its running integral:
\begin{equation} \label{eq:radialintegral}
    Q_{ind}(r) = \int_{0}^{r} \sigma_{ind}(r') 2\pi r' \,{\rm d}r'
    \; .
\end{equation}

%%%%%%%%%%%%%%%%%%%%%%%   %%%%%%%%%%%%%%%%%%%%%%%

\section{Charge induced inside the metal}
\label{sec:results:inducedq}

We first focus on the influence of both the metallicity and the nature of the solvent on the charge induced in the metal by an interfacial ion. In addition, we assess the ability of MDFT to accurately predict the induced charge by comparing the results obtained through functional minimization with those of MD simulations. The structure of the solvent in all cases will be discussed in detail in Section~\ref{sec:results:solvstr}.

%%%%%%%%%%%%%%%%%%%%%%%   
\subsection{Two-dimensional distribution of the induced charge}
\label{sec:results:inducedq:2d}

\subsubsection{Sodium ion at the water/graphite interface}

\begin{figure*}[!ht]
\center
\includegraphics[width=5in]{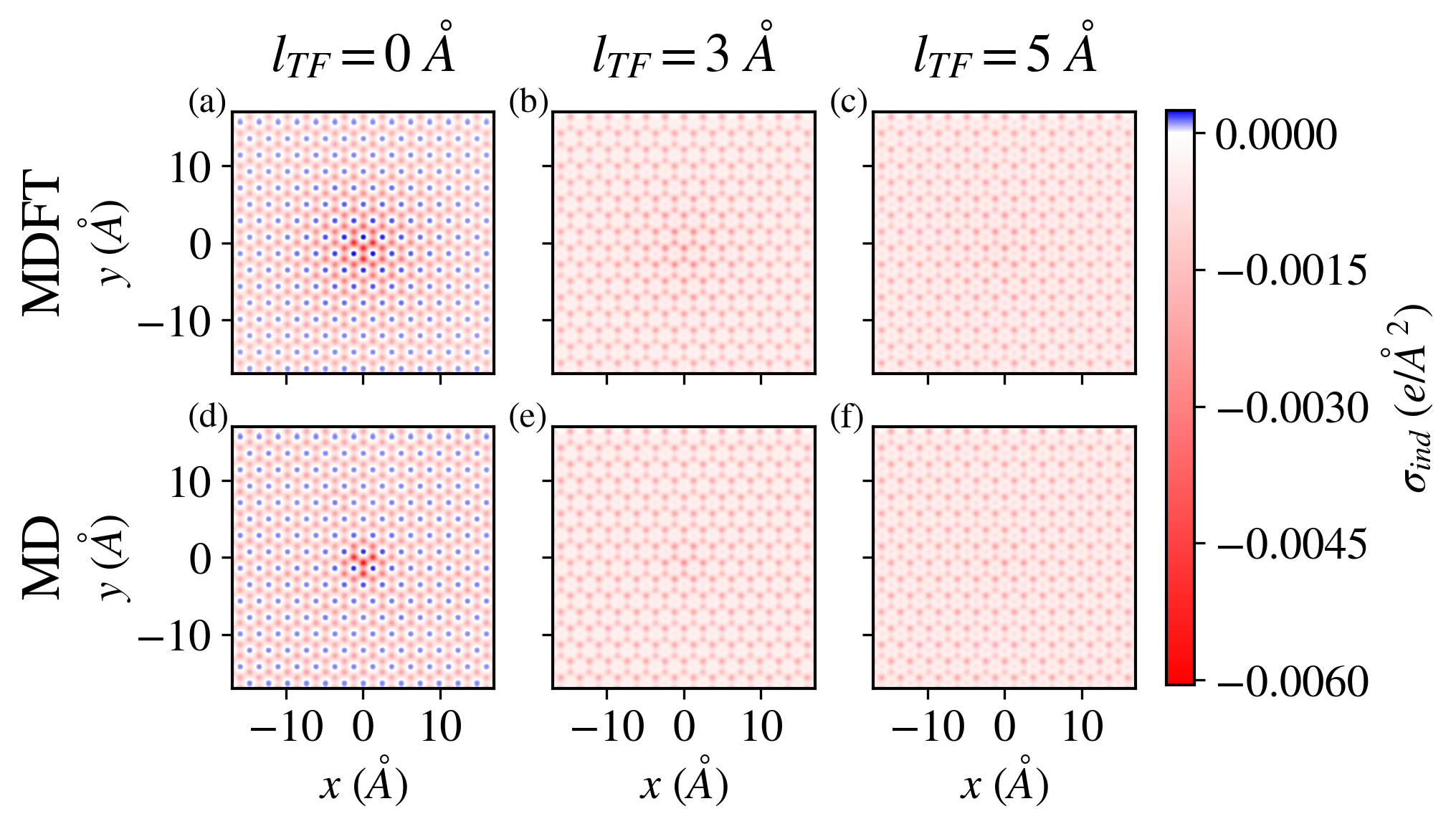}
\caption{Surface charge density $\sigma_{ind}(x,y)$ (see Eq.~\ref{eq:2d}) induced by a Na$^+$ cation in liquid water located at $(x,y)=(0,0)$ and a distance $z_{ion}=5.1$~\AA\ from the first atomic plane of the electrode, for $l_\text{TF}=0, 3$ and $5$~\AA. Panels (a)-(c) indicate the results from MDFT, panels (d)-(f) the ones from MD. The colormap is common for all panels.}
\label{fig2D:water}
\end{figure*}

Starting with the aqueous system depicted in Fig.~\ref{fig:snapshot}a, we show in Fig.~\ref{fig2D:water} the surface charge density $\sigma_\text{ind}(x,y)$ (see Eq.~\ref{eq:2d:MD}) induced by a Na$^+$ ion located at $(x,y)=(0,0)$ and a distance $z_{ion}=5.1$~\AA\, from the first atomic plane of the electrode, for three values of the Thomas-Fermi screening: $l_\text{TF}=0, 3$ and $5$~\AA. The MDFT predictions (panels a-c) are in excellent agreement with the MD results (panels d-f). In all cases, the atomic structure of graphite is clearly visible in the modulation of the induced charge density. 
While the metal is overall negatively charged (its total charge compensates that of the Na$^+$ ion), a pattern of slightly positive local charges (in blue) is observed for $l_\text{TF} = 0$~\AA. As already discussed in the case of gold~\cite{pireddu_molecular_2021}, these regions correspond to atoms in the second atomic layer, with a slightly positive charge induced by the neighboring negative induced charge in the first atomic plane in contact with the liquid. The charge in this first plane is more negative below the ion and decays away from it, as discussed in more detail with the radial charge distribution in Section~\ref{sec:results:inducedq:r}. As the screening length increases, the lateral decay become slower and leads to a homogeneous distribution. This behavior parallels observations made in previous studies, where increasing $l_\text{TF}$ led to broader charge distributions~\cite{Nair-2024} and reduced charge localization, attributed to the energetic cost of high local charge densities~\cite{scalfi-pnas-2021}.

%%%%%%%%%%%%%%%%%%%%%%%
\subsubsection{Sodium ion at the acetonitrile/graphite interface}

\begin{figure*}[!ht]
\center
\includegraphics[width=5in]{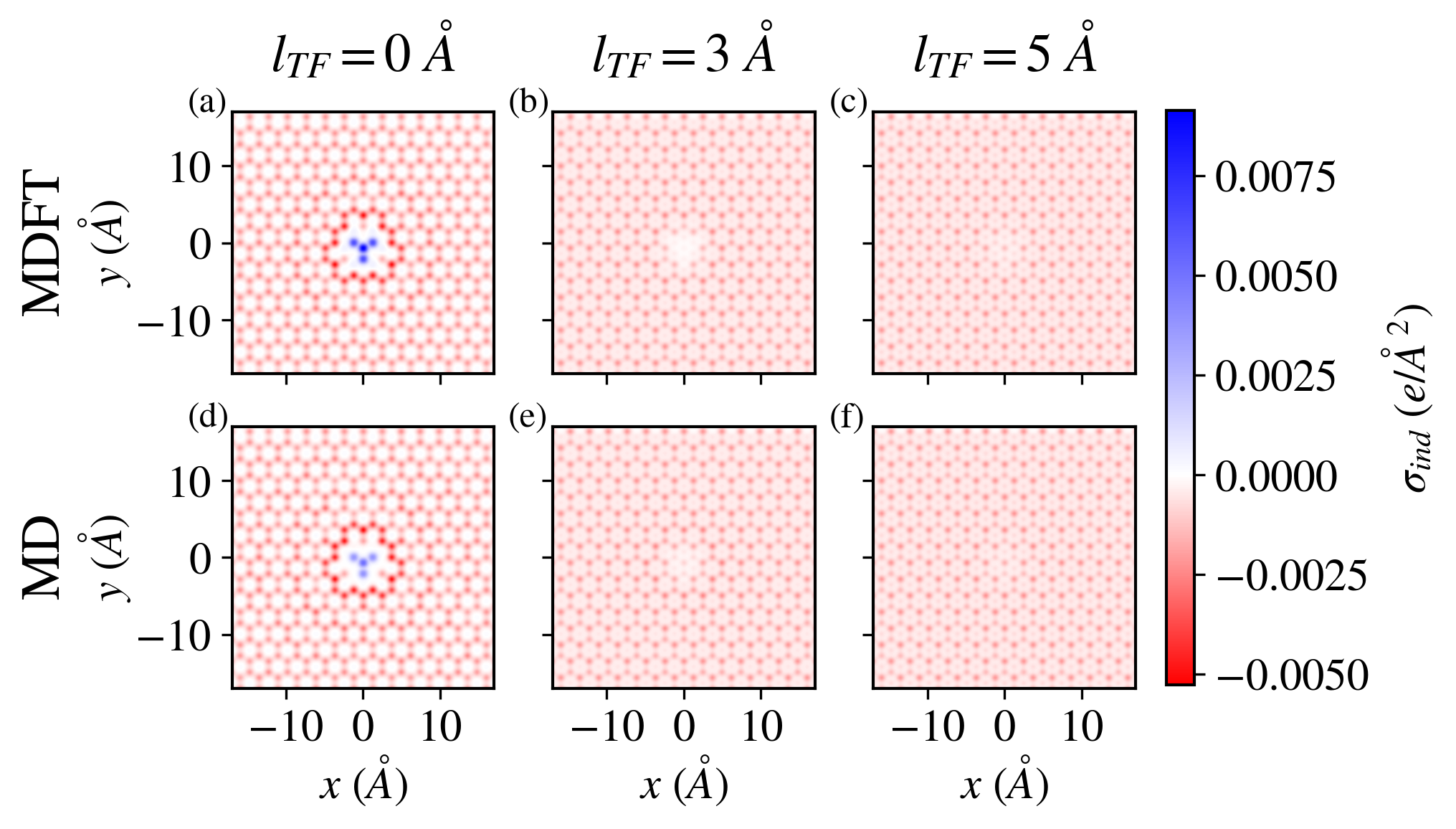}
\caption{
Surface charge density $\sigma_{ind}(x,y)$ (see Eq.~\ref{eq:2d}) induced by a Na$^+$ cation in liquid acetonitrile located at $(x,y)=(0,0)$ and a distance $z_{ion}=5.1$~\AA\ from the first atomic plane of the electrode, for $l_\text{TF}=0, 3$ and $5$~\AA. Panels (a)-(c) indicate the results from MDFT, panels (d)-(f) the ones from MD. The colormap is common for all panels.}
\label{fig2D:Mecn}
\end{figure*}

We now consider a Na$^+$ ion dissolved in acetonitrile, at the same position as in the water case. Fig~\ref{fig2D:Mecn} shows the surface charge density for $l_\text{TF}=0, 3$ and $5$~\AA. In contrast to the water case (note in particular the different range of $\sigma_\text{ind}$ compared to Fig.~\ref{fig2D:water}), for $l_\text{TF} = 0$~\AA, the acetonitrile system exhibits a positive charge in the first electrode plane below the ion, as shown by the blue color in panels (a) and (d), surrounded by a localized negative charge, followed by a smooth decay away from the ion. As will be discussed in Section~\ref{sec:results:solvstr}, this reflects the presence of acetonitrile molecules with electronegative nitrogen atoms solvating the cation. This subtle effect is well reproduced by MDFT, even though the local charge is more positive below the ion than in the MD case. MDFT also captures well the effect of screening inside the metal, which leads to a more delocalized charge distribution (in particular there is no sign change, unlike for $l_\text{TF} = 0$~\AA), even though the agreement with MD is not quantitative. In particular, the ability of MDFT to accurately capture the complex pattern of the induced charge density indicates a proper description of the solvation structure (see Section~\ref{sec:results:solvstr:ionsolv}).

%%%%%%%%%%%%%%%%%%%%%%%  
\subsection{Radial distribution of the induced charge}
\label{sec:results:inducedq:r}

%%%%%%%%%
\subsubsection{Sodium ion in water and acetonitrile}

\begin{figure}[!ht]
\center
\includegraphics[width=3.4in]{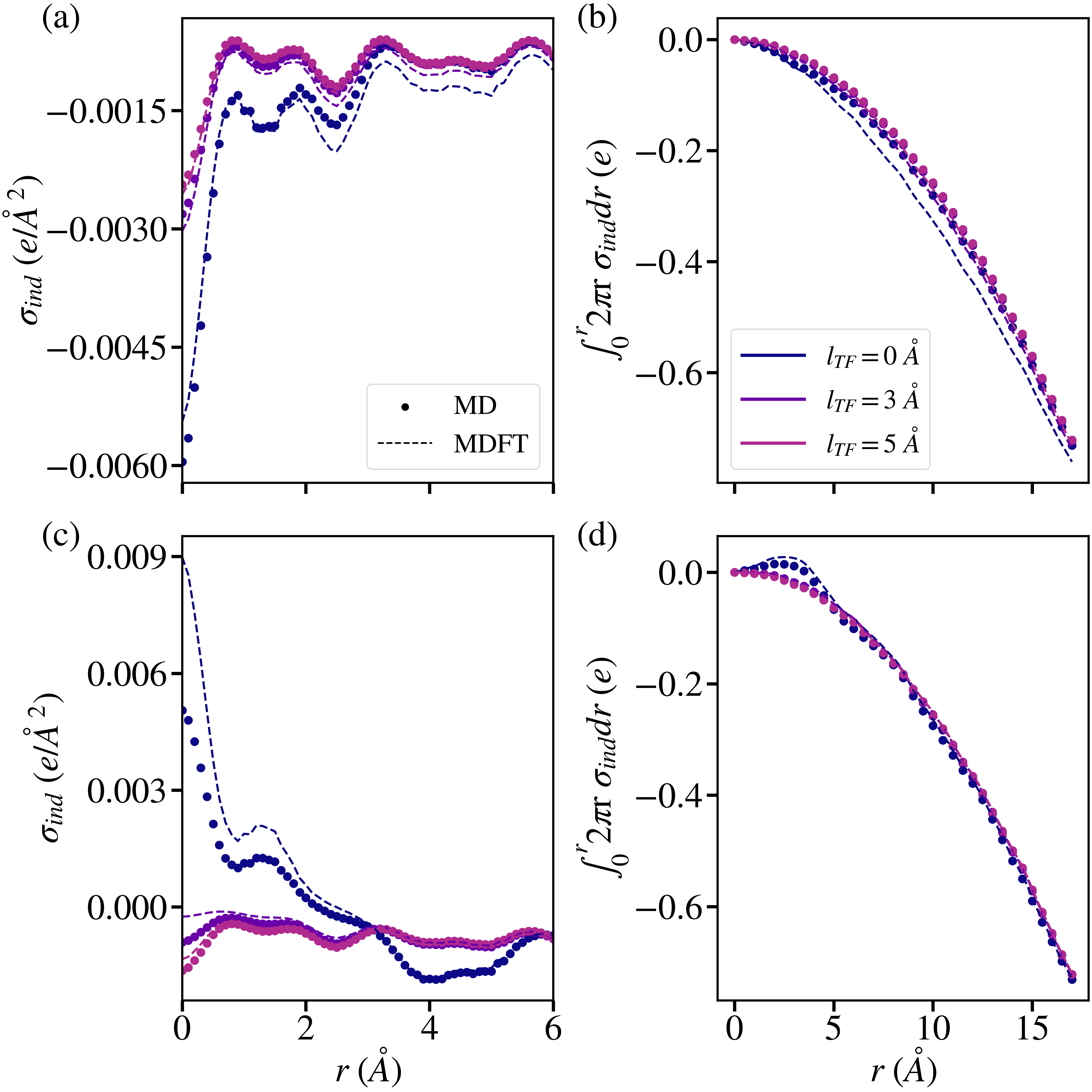}
\caption{ (a, c) Radially averaged induced surface charge density $\sigma_{ind}(r)$ (see Eq.~\ref{eq:rad}), and (b, d) the corresponding running integrals (see Eq.~\ref{eq:radialintegral}) for the charge density induced in a graphite electrode by a Na$^+$ cation located at $(x, y) = (0, 0)$ and positioned $z_{ion}=5.1$~\AA\, above the electrode’s first atomic plane. Colors refer to Thomas-Fermi lengths $l_\text{TF}=0, 3$ and $5$~\AA\, as indicated in panel (b). Panels (a) and (b) refer to Na$^+$ in water, while (c) and (d) refer to Na$^+$ in acetonitrile.}
\label{fig:rad}
\end{figure}

In order to facilitate the discussion of the effect of the Thomas-Fermi screening length and the comparison between MDFT and MD, we now turn to the radial induced charge density $\sigma_{ind}(r)$ and its running integral $Q_{ind}(r)$ introduced in Eqs.\ref{eq:rad} and \ref{eq:radialintegral} respectively. The results are shown in Fig.~\ref{fig:rad}, for Na$^+$ in water in panels~\ref{fig:rad}a and~\ref{fig:rad}b, and in acetonitrile in panels~\ref{fig:rad}c and~\ref{fig:rad}d. In each panel, the color indicates the value of $l_\text{TF}$ and MD results (symbols) are compared with the MDFT ones (dashed lines). As discussed previously in the case of vacuum~\cite{Nair-2024}, the oscillations observed in $\sigma_{ind}(r)$ with both solvents (panels~\ref{fig:rad}a and~\ref{fig:rad}c) arise from the atomic structure of the electrode. Their magnitude reflects the influence of the solvent. The ability of MDFT to capture these oscillations further establishes its suitability to describe the solvent implictly in realistic atomistic models (ion and explicit electrode atoms). These oscillations are smoothed out in the running integral, $Q_{ind}(r)$, which does not converge to $-q_{ion}$ (which corresponds to the whole electrode) but rather to a value close to $-\frac{\pi}{4}q_{ion}$ because the corners of the electrode are not included when integrating radially up to half of the simulation box. Overall, MDFT is able to reproduce the MD ones, including both the effect of $l_\text{TF}$ and that of the solvent (water vs acetonitrile). 

A closer examination of the water case (panels~\ref{fig:rad}a and~\ref{fig:rad}b) reveals a small discrepancy for $l_\text{TF}= 0$~\AA. This is probably due to subtle differences in the solvent structure predicted by MD and MDFT, mainly in the first adsorbed layer at the graphite interface and in the first solvation shell around the sodium ion. This is a known limitation of the excess functional, which underestimates the tetrahedral order around charged solutes. As $l_\text{TF}$ increases, the local charge below the ion ($r=0$) decreases, while the lateral extent of the charge distribution broadens and becomes more homogeneous due to the combined effects of the screening inside the metal and the presence of the solvent. 

The results for acetonitrile (panels~\ref{fig:rad}c and~\ref{fig:rad}d) are similar. As already observed in the 2D maps of Fig.~\ref{fig2D:Mecn}a and~\ref{fig2D:Mecn}, for $l_\text{TF}$ = 0~\AA\ $\sigma_{ind}(r)$ displays a positive region near $r=0$. In that case, MDFT overestimates $\sigma_{ind}(r)$ compared to MD, in line with previous findings that MDFT tends to overestimate solvation structures relative to MD simulations~\cite{zhao-2011}. As $l_\text{TF}$ increases, the local charge below the ion ($r=0$) decreases, while the lateral extent of the charge distribution broadens, improving the agreement between MDFT and MD results. Interestingly, for larger $r$ values, $l_\text{TF}$ exhibits a reversed ordering compared to the water solvent. For the radial integral $Q_{ind}(r)$, the agreement between MDFT and MD remains quantitative.

\begin{figure}[!ht]
\center
\includegraphics[width=3.4in]{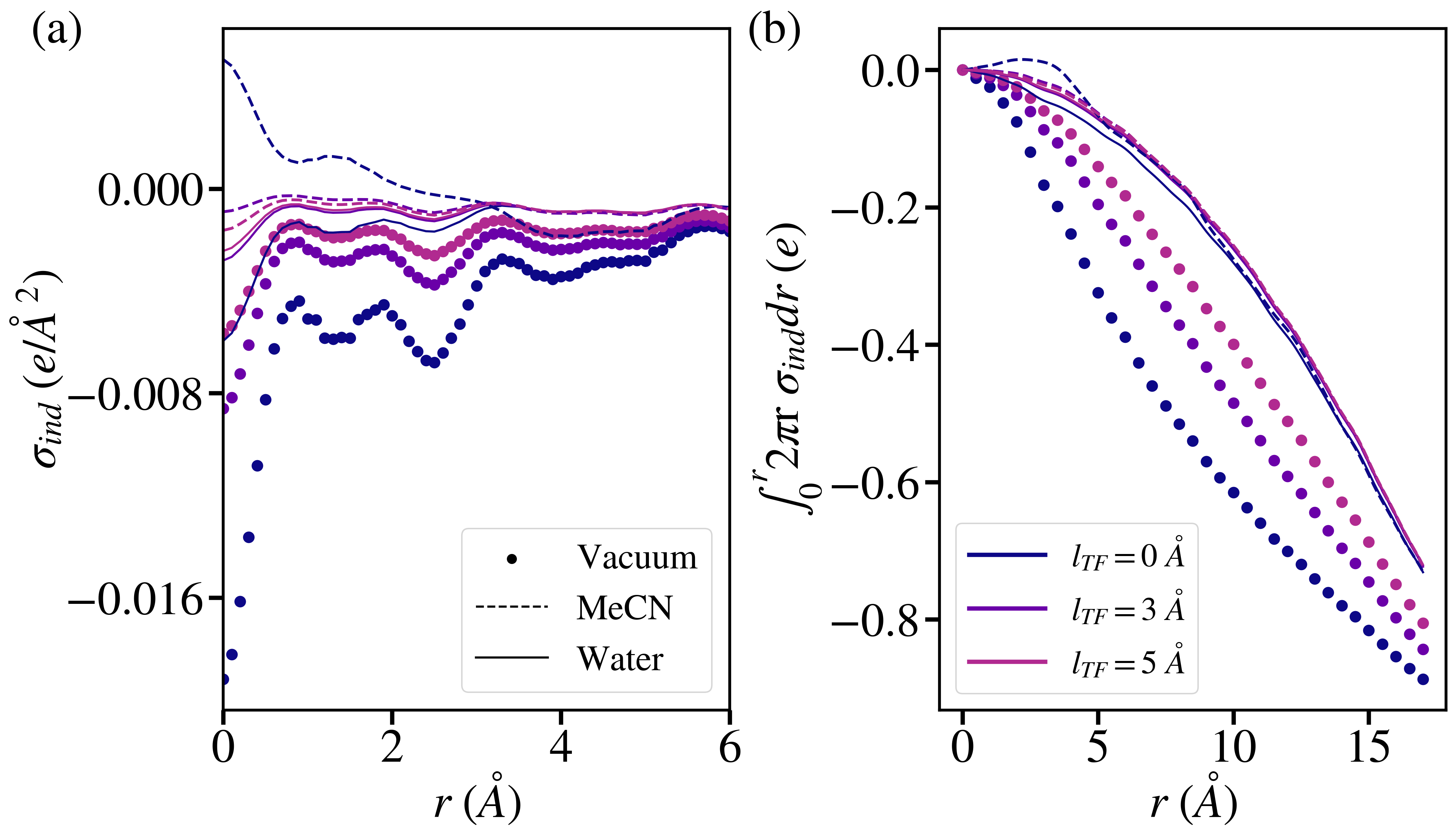}
\caption{ (a) Radially averaged induced surface charge density $\sigma_{ind}(r)$ (see Eq.~\ref{eq:rad}), and (b) the corresponding running integrals (see Eq.~\ref{eq:radialintegral}) for the charge density induced in a graphite electrode by a Na$^+$ cation located at $(x, y) = (0, 0)$ and positioned $z_{ion}=5.1$~\AA\, above the electrode’s first atomic plane. Symbols are the results in the absence of solvent, while dashed and solid lines correspond to acetonitrile and water, respectively. Colors refer to Thomas-Fermi lengths $l_\text{TF}=0, 3$ and $5$~\AA\, as indicated in panel (b).}
\label{fig:radwithvacuum}
\end{figure}

As mentioned in the introduction, Kornyshev and co-workers~\cite{vorotyntsev_1980} predicted using continuum electrostatics that the characteristic length for the long-range lateral decay of the charge/potential induced in the metal by an ion scales as $l_\text{TF}\epsilon_\textrm{solv}/\epsilon_\textrm{met}$. The MD results in the presence of water and acetonitrile (already shown in Fig.~\ref{fig:rad}) are compared to the vacuum case, discussed in more detail in Ref.~\citenum{Nair-2024}, in Fig.~\ref{fig:radwithvacuum}. The influence of $l_\text{TF}$ is much more visible (in particular in the running integral of panel~\ref{fig:radwithvacuum}b) in vacuum than in the presence of a solvent. For both solvents, the broadening of the distribution with respect to the vacuum case, consistent with the increase predicted from continuum electrostatics, is in fact such that the induced charge density is almost homogeneous beyond a few~\AA\ from the ion, regardless of $l_\text{TF}$. Such a homogeneity (also visible in Figs.~\ref{fig2D:water} and~\ref{fig2D:Mecn}) is due to the large decay length compared to the lateral dimensions of the box (see Ref.~\citenum{Nair-2024} for more discussion of the effect of periodic boundary conditions in the vacuum case) and prevents us from a quantitative test of the scaling as $l_\text{TF}\epsilon_\textrm{solv}/\epsilon_\textrm{met}$. We note that this predicts no effect of the solvent permittivity in the $l_\text{TF}=0$ case, unlike what is observed in simulations. However this prediction is only valid for a single ion in an infinite system, in the limit of large distances from the ion, so that it is not in contradiction with the MD result that are sensitive to the periodic boundary conditions in the lateral directions.

%%%%%%%%%
\subsubsection{Chloride ion in water and acetonitrile}

\begin{figure}[!ht]
\center
\includegraphics[width=3.4in]{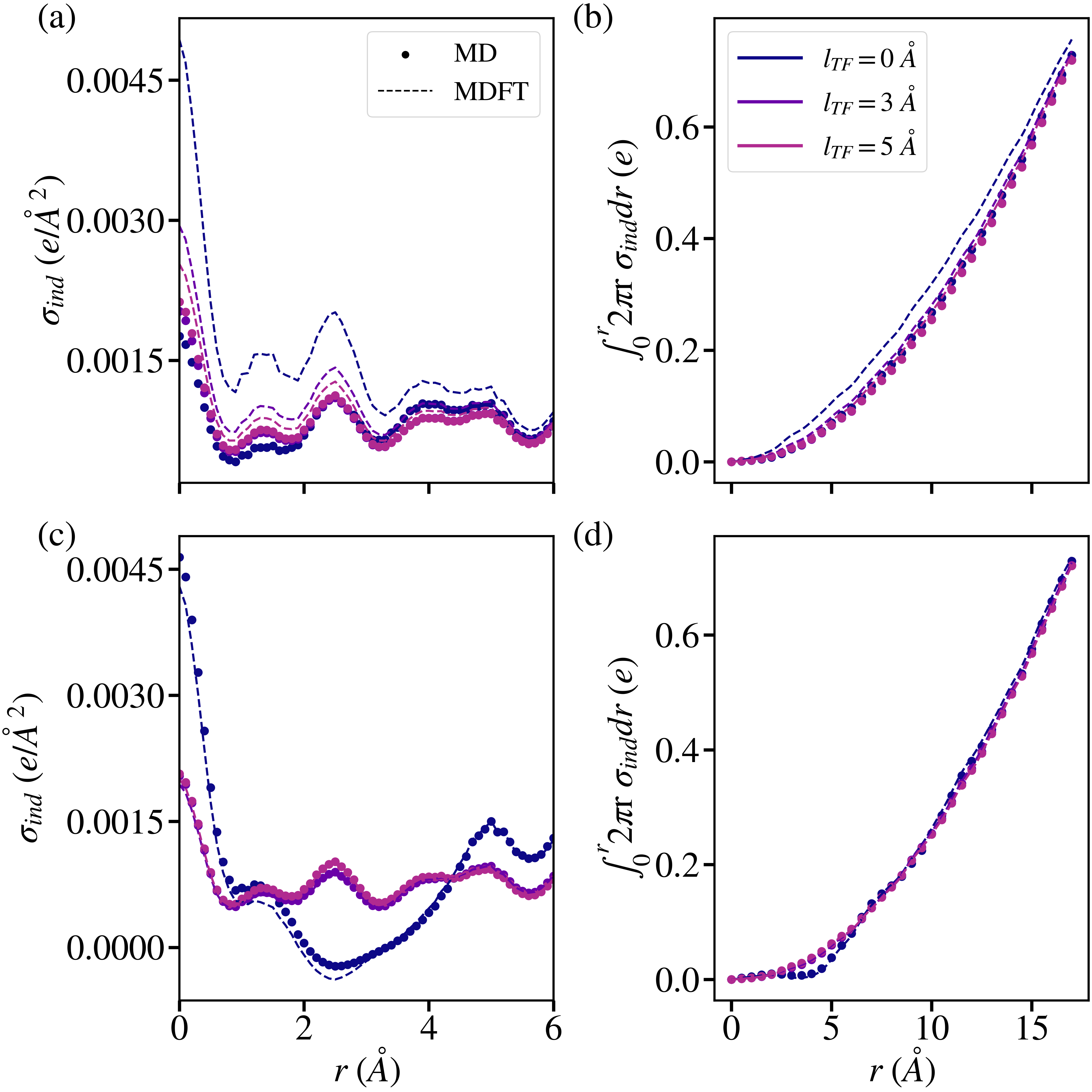}%rprofile_Cl.png}
\caption{(a, c) Radially averaged induced surface charge density $\sigma_{ind}(r)$ (see Eq.~\ref{eq:rad}), and (b, d) the corresponding running integrals (see Eq.~\ref{eq:radialintegral}) for the charge density induced in a graphite electrode by a Cl$^-$ anion located at $(x, y) = (0, 0)$ and positioned $z_{ion}=5.1$~\AA\, above the electrode’s first atomic plane. Colors refer to Thomas-Fermi lengths $l_\text{TF}=0, 3$ and $5$~\AA\, as indicated in panel (b). Panels (a) and (b) refer to Cl$^-$ in water, while (c) and (d) refer to Cl$^-$ in acetonitrile.}
\label{fig:rad-cl}
\end{figure}

To further investigate MDFT’s predictive capabilities, we replace Na$^+$ with Cl$^-$ at the same position in both solvents. While the 2D distributions of the induced charge are similar for MDFT and MD (see Section.~S1 of the supplementary material), the radial distributions $\sigma_{ind}(r)$ indicate some quantitative deviations, most notably for water (see Figs.~\ref{fig:rad-cl}a and ~\ref{fig:rad-cl}b). This is another consequence of the limitations of the excess functional, which is known to perform better for positively charged solutes than for negatively charged ones, due to its shortcomings for the description of hydrogen bonding and to the orientation of OH bonds toward the anion. In acetonitrile, where no such complex bonding interactions occur, the functional is more accurate, and we observe excellent agreement between MD and MDFT for both the radial distribution and the its running integral (see Figs.~\ref{fig:rad-cl}c and~\ref{fig:rad-cl}d).

%%%%%%%%%%%%%%%%%%%%%%%   %%%%%%%%%%%%%%%%%%%%%%%
\section{Solvent structure}
\label{sec:results:solvstr}

Overall, the above results confirm both the impact of metallicity on the charge induced inside the metal and the consistency between MDFT and MD for this observable. In this section, we explore how the Thomas-Fermi screening length influences the structure of the solvent at the interface with graphite, first in the absence of ion, then investigating the solvation of cations and anions at the interface.

%%%%%%%%%%%%%%%%%%%%%%%
\subsection{Interface between graphite and pure solvents}
\label{sec:results:solvstr:pure}

%%%%%%%%%%%%%%%%%%%%%%%
\subsubsection{Solvent layering}
\label{sec:results:solvstr:pure:layering}

\begin{figure}[!ht]
\center
\includegraphics[width=3.5in]{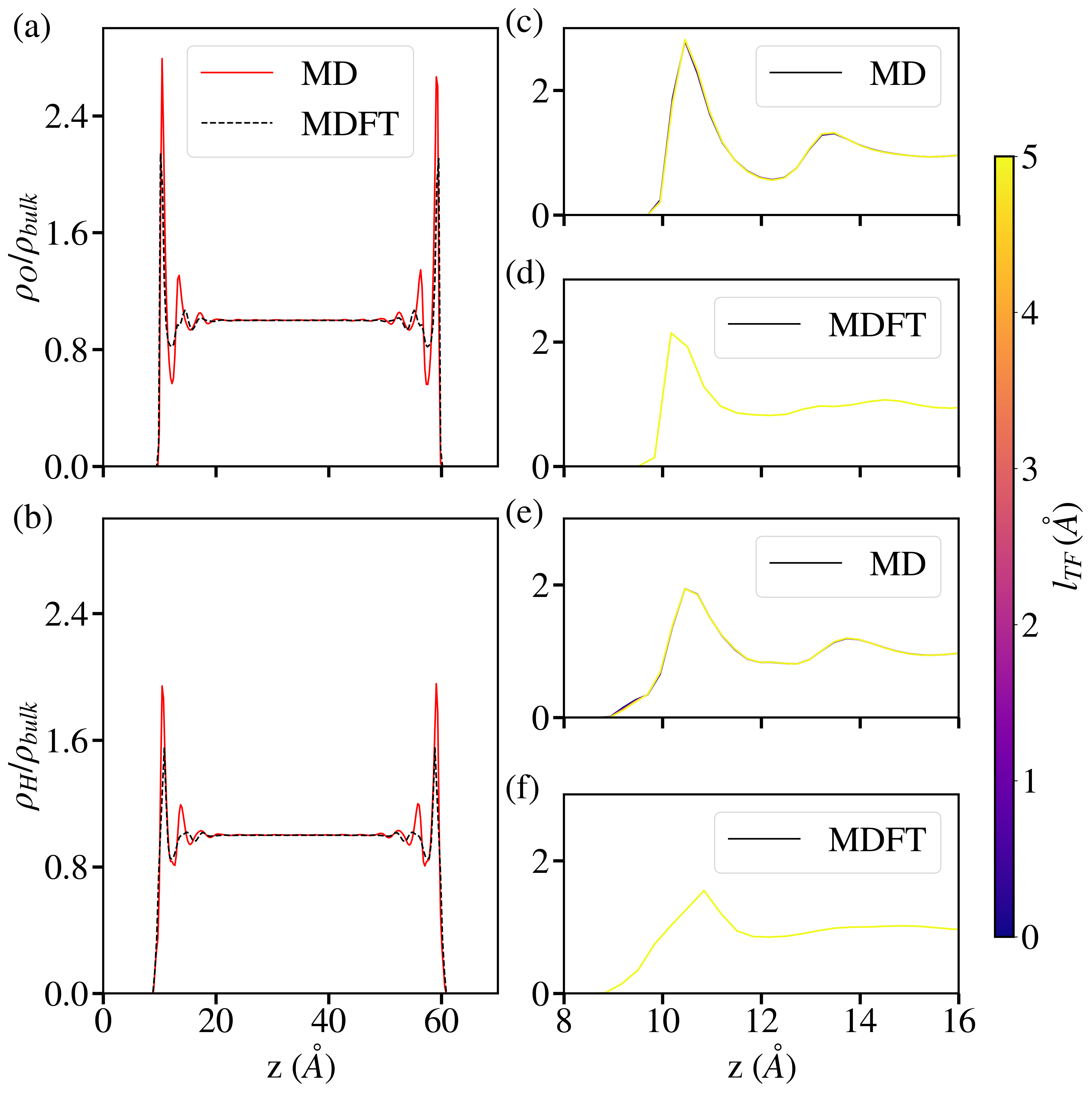}
\caption{
Oxygen (a) and hydrogen (b) density profiles at the graphite/water interface for $l_\text{TF}$ =0~\AA. The MD results are shown in red, while the MDFT results are represented by black dashed lines. Panels (c) and (d) compare the oxygen density profiles for different Thomas-Fermi lengths $l_\text{TF}$ (indicated by the color), for MD and MDFT, respectively, while hydrogen density profiles are shown in panels (e) and (f) for MD and MDFT, respectively. The profiles are zoomed in for clarity.}
\label{fig:den}
\end{figure}

Fig.~\ref{fig:den} presents a comparison of water's oxygen  ($\rho_\text{O}$) and hydrogen ($\rho_\text{H}$) density profiles computed using both MD and MDFT. The two methods yield qualitatively similar results, with two distinct molecular layers, beyond which the homogeneous densities are rapidly recovered. However, a closer examination reveals some differences. For the oxygen density in  Fig.~\ref{fig:den}a, the two main peaks of the density profile are smaller in MDFT compared to MD. In  Fig.~\ref{fig:den}b, the maximum of the first hydrogen density peak is located at 10.5~\AA\ in MDFT, slightly closer to the electrode than in MD (10.8~\AA). In addition, this first peak is sharper in MDFT than in MD, consistently with previous findings~\cite{Jeanmairet-2019}. 

Furthermore, we compare the oxygen (Figs.~\ref{fig:den}c and~\ref{fig:den}d) and hydrogen (Figs.~\ref{fig:den}e and~\ref{fig:den}f) density profiles for various values of $l_\text{TF}$ ranging from 0 to 5~\AA\, as indicated in the color bar. The results are shown for both MD and MDFT. We observe that the solvent density remains largely unchanged with varying $l_\text{TF}$ values. This aligns closely with the observations of Scalfi \emph{et al.}~\cite{Scalfi_tfmodel_2020}, where water density near the electrode showed minimal variation with different $l_\text{TF}$ values in the absence of applied voltage across the interface. This  suggest that, under these conditions, the metallicity of the electrode has a limited impact on the structure of the solvent at the interface. This is further supported by the fact that the O and H densities on the neutral wall (which corresponds to the insulating limit $l_\text{TF}\to\infty$), visible on the right sides of panels Figs.~\ref{fig:den}a and~\ref{fig:den}b, are comparable to that on the metallic one (only symmetric with respect to the center of the pore). This conclusion might not hold for non-zero applied voltage~\cite{Scalfi_tfmodel_2020}. 

\begin{figure}[!ht]
\center
\includegraphics[width=3in]{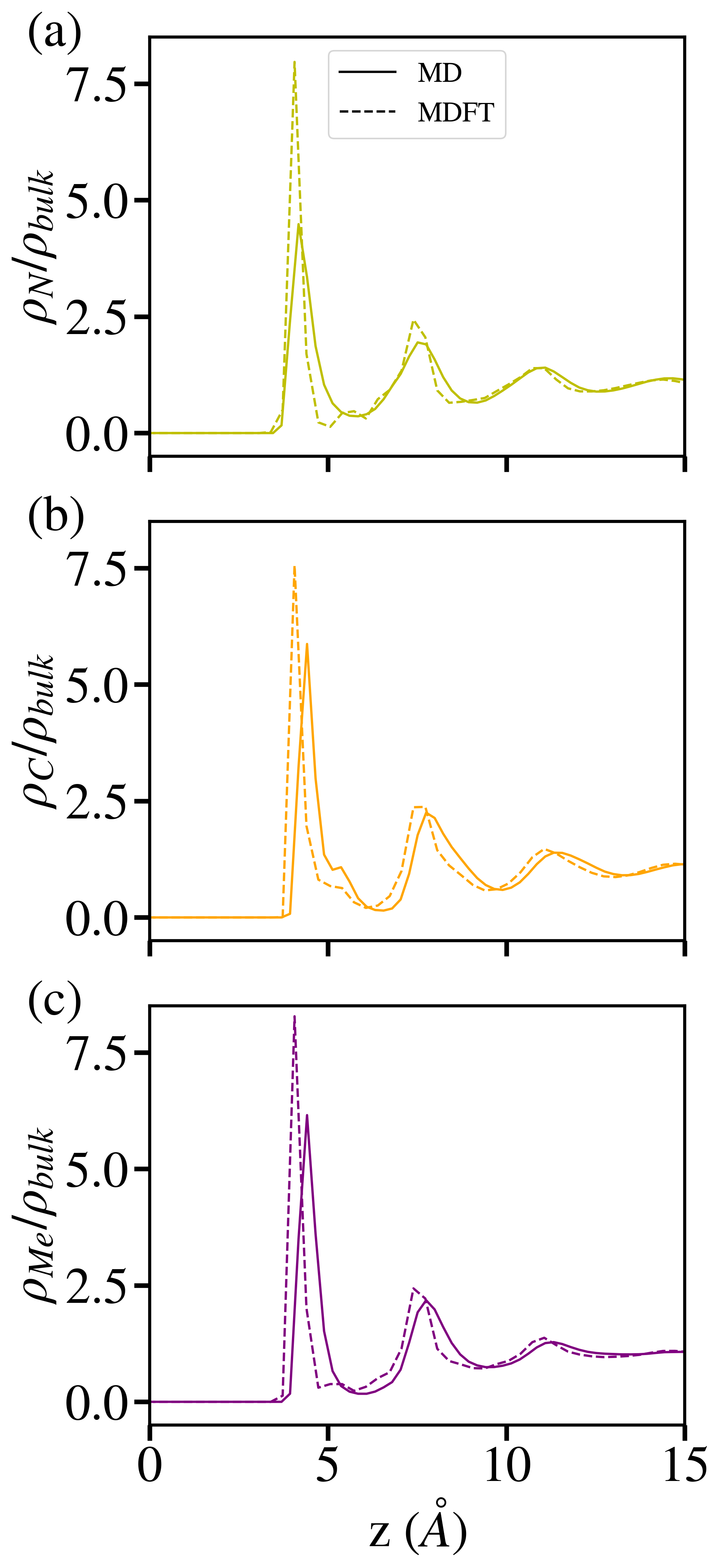}
\caption{
Nitrogen (a), carbon (b), and methyl group (c) density profiles at the acetonitrile/graphite interface for $l_\text{TF}$ = 0~\AA, as obtained by  MDFT (black dashed lines) and MD (blue solid lines). }
\label{fig:denmecn}
\end{figure}

We also consider the predictions by MDFT and MD of the nitrogen (Fig.~\ref{fig:denmecn}a), carbon (Fig.~\ref{fig:denmecn}b), and methyl group (Fig.~\ref{fig:denmecn}c) density profiles at the acetonitrile/graphite interface for $l_\text{TF}$ = 0~\AA. Both MDFT and MD predict an interfacial structure with 3-4 distinct layers, with density maxima of decreasing intensity away from the interface. However, closer examination reveals small differences between the predictions of the two methods. For all profiles, the first solvation peak predicted by MDFT is positioned closer to the electrode than with MD.  Notably, the carbon and methyl group profiles reveal significant differences between MDFT and MD, with peak maxima at 4.06~\AA\ for MDFT and 4.36~\AA\ for MD. For nitrogen, the difference is smaller, with peak maxima at 4.06~\AA\ for MDFT and 4.16~\AA\ for MD. Much like water, the solvent structure remains largely unaffected by changes in the Thomas-Fermi length (not shown). 
%We have also examined the effect of varying $l_\text{TF}$ on the density profiles of MeCN, and similar to the case of water, no significant change is observed. This further confirms that the solvent structure remains largely unaffected by variations in $l_\text{TF}$ under the conditions studied. %\tocheck{Figure is not added.}

%%%%%%%%%%%%%%%%%%%%%%%
\subsubsection{Solvent structure in the first adsorbed layer on graphite}
\label{sec:results:solvstr:pure:1stlayer}

\begin{figure}[!ht]
\center
\includegraphics[width=3.5in]{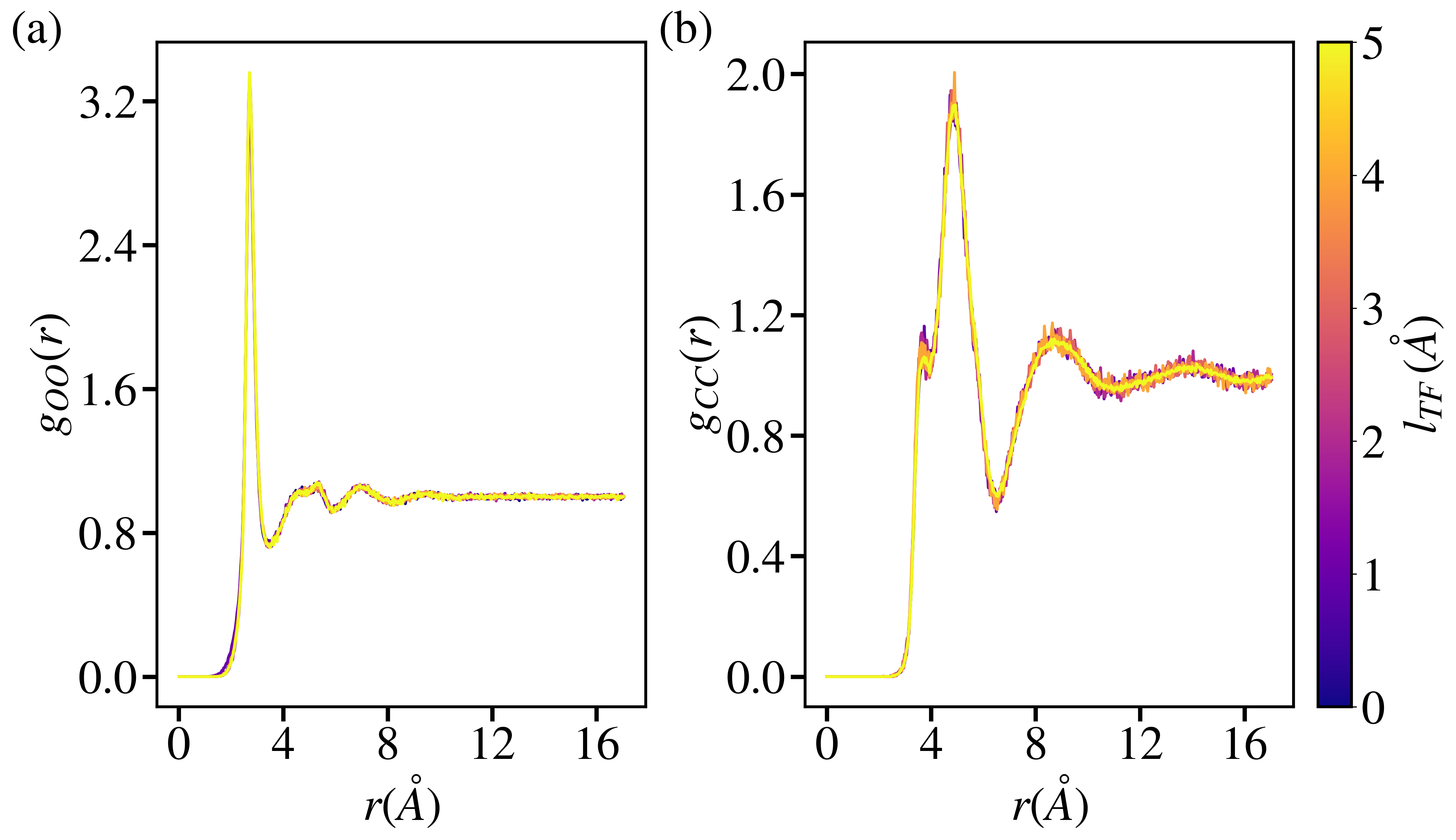}
\caption{
Lateral correlations between solvent molecules in the first adsorbed layer on the graphite electrode, for (a) water, with the 2D radial distribution for oxygen atoms, and (b) acetonitrile, with the 2D radial distribution for the central carbon atoms, for a range of $l_\text{TF}$ indicated by the color bar.
}
\label{fig:gooandgcc}
\end{figure}

Since the electrode metallicity has a clear impact on the surface charge density within the electrode, as discussed in Section~\ref{sec:results:inducedq}, whereas it virtually has no effect on the density profiles in the direction perpendicular to the surface, we now examine its possible influence on the 2D structure of the solvent within the first adsorbed layer (defined from the density profiles of Figs.~\ref{fig:den} and~\ref{fig:denmecn}). The 2D radial distribution functions between oxygens of water molecules and between central carbon atoms of acetonitrile molecules within that layer are shown presented in Figs.~\ref{fig:gooandgcc}a and Figs.~\ref{fig:gooandgcc}b, respectively, for Thomas-Fermi screening lengths varying from 0 to 5~\AA. There is a clear in-plane structure, with a first peak in the radial distribution functions located at $2.7$~\AA\ for water and $4.8$~\AA\ for acetonitrile. These distances are not due to a templating effect due to the graphite lattice (the 2D densities along the plane do not display any particular feature, not shown) and rather reflect the lateral correlations between molecules sliding over the surface. While the presence of the wall does not change the typical distance between water molecules compared to the bulk, it results in an increase of the linear acetonitrile molecules ($\approx 4$~\AA\ in the bulk) due to packing constraints. Importantly, we observe no significant effect of the Thomas-Fermi length on this in-plane solvent structure, in contrast with that of the induced charge. This is consistent with previous findings on a nanocapacitor consisting of a 1~M aqueous NaCl solution between graphite electrodes~\cite{scalfi-pnas-2021}.

%%%%%%%%%%%%%%%%%% %%%%%%%%%%%%%%%%%%%%%%%%
%\subsection{Average solvent density in the (r,z) plane : MDFT vs. MD}
\subsection{Ion solvation at the solvent/graphite interface}
\label{sec:results:solvstr:ionsolv}

We now turn to the solvation of ions by water and acetonitrile at the interface with graphite, by considering the prototypical cases sodium cation and chloride anion. The effect of the ionic charge is twofold: Firstly, it structures the solvent in its vicinity and secondly, it induces a non-zero charge on the metallic electrode, which may in turn modify the structure of the interfacial solvent. The induced charge was described in Section~\ref{sec:results:inducedq} and we consider here only the solvent structure. To that end, we introduce the radially average solvent charge density in planes perpendicular to the electrode containing the ion:
\begin{equation}
\label{eq:rhosolv}
    \rho_\text{solv}(r,z) = \left \langle \sum_k q_k\delta(r_k-r)\delta(z_k-z) \right \rangle \; ,
\end{equation}
where the sum runs over solvent atoms with partial charge $q_k$ and position $(r_k,z_k)$ expressed in cylindrical coordinates. We first examine the effect of the ion charge for $l_\text{TF} = 0$~\AA, before turning to the effect of $l_\text{TF}$, and compare MDFT with MD in all cases.

%%%%%%%%%%%%%%%%%%
\subsubsection{Effect of ion charge}

\begin{figure*}[!ht]
\center
\includegraphics[width=6in]{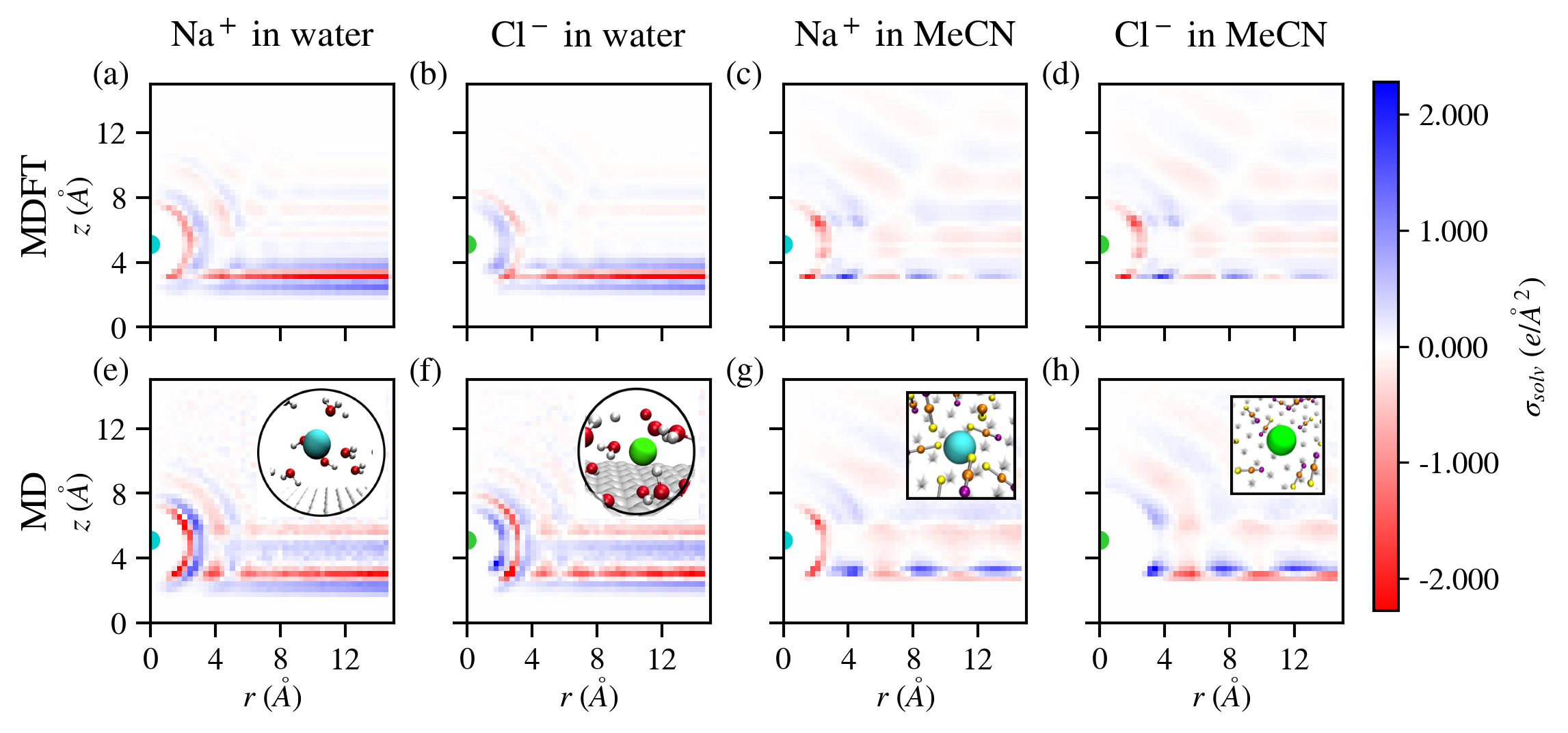} 
\caption{Average charge density distribution (see Eq.~\ref{eq:rhosolv}) of solvent molecules around Na$^+$ and Cl$^-$ ions in water (a,b,e,f) and acetonitrile (c,d,g,h) for $l_\text{TF} = 0$~\AA\ obtained using MDFT (a-d) and MD (e-h). Insets illustrate typical configurations of solvent molecules in the first solvation shell, with water represented using red for oxygen and white for hydrogen, and acetonitrile using yellow for nitrogen, orange for carbon, and purple for the methyl group.}
\label{fig:rzmap}
\end{figure*}

Fig.~\ref{fig:rzmap} shows the average solvent charge density in the $(r,z)$ plane for Na$^+$ and Cl$^-$ ions in water and acetonitrile. The alternating signs of the charge density in panels~\ref{fig:rzmap}a, \ref{fig:rzmap}b, \ref{fig:rzmap}e and~\ref{fig:rzmap}f reflect the characteristic orientation of water molecules around the ions and at the graphite surface. Far from the ions, the charge density profiles depend only on $z$ and are consistent with the atomic density profiles of Fig.~\ref{fig:den}. Close to the ion, water molecules are oriented with their oxygen atoms pointing toward the sodium cation, while the preferred orientation is flipped in the vicinity of the chloride atom, as expected. MDFT is able to reproduce faithfully the rather complex solvation behaviour for both ions, although minor differences arise in the magnitude of the charge densities.

Similar conclusions can be drawn in the acetonitrile case, shown in panels~\ref{fig:rzmap}c, \ref{fig:rzmap}d, \ref{fig:rzmap}g and~\ref{fig:rzmap}h. Far from the ions, we do not observe the alternating positively and negatively charged planes perpendicular to the electrode. Instead, we observe an oscillating charge within the first adsorbed layer and a more diffuse charge pattern beyond, suggesting the packing of the linear molecules parallel to the surface. The nitrogen atom is positioned closer to the Na$^+$ cation, while it is the methyl groups which points towards Cl$^-$. This can be attributed to acetonitrile's linear structure and the higher electronegativity of nitrogen (as described within this classical force field by its negative partial charge) compared to the carbon atom and methyl group. MDFT effectively captures these asymmetric solvation behaviors resulting with an overall good agreement with MD.

Finally, It is worth noticing that the noisy patterns far from the ion and the electrode in MD is absent in MDFT, which predicts a well defined charge density in the whole $(r,z)$ plane. Indeed, the solvent density predicted by MDFT is obtained by minimizing the functional, while the one predicted by MD suffer from statistical error (which can be reduced using force-based estimators~\cite{borgis_computation_2013,coles_computing_2019,rotenberg_use_2020,coles_reduced_2021}). This is particularly true in region where there is no preferred solvent orientation, which is the case far from the solute and the electrode.

%%%%%%%%%%%%%%%%%% 
\subsubsection{Effect of Thomas-Fermi screening length}

\begin{figure*}[!ht]
\center
\includegraphics[width=6in]{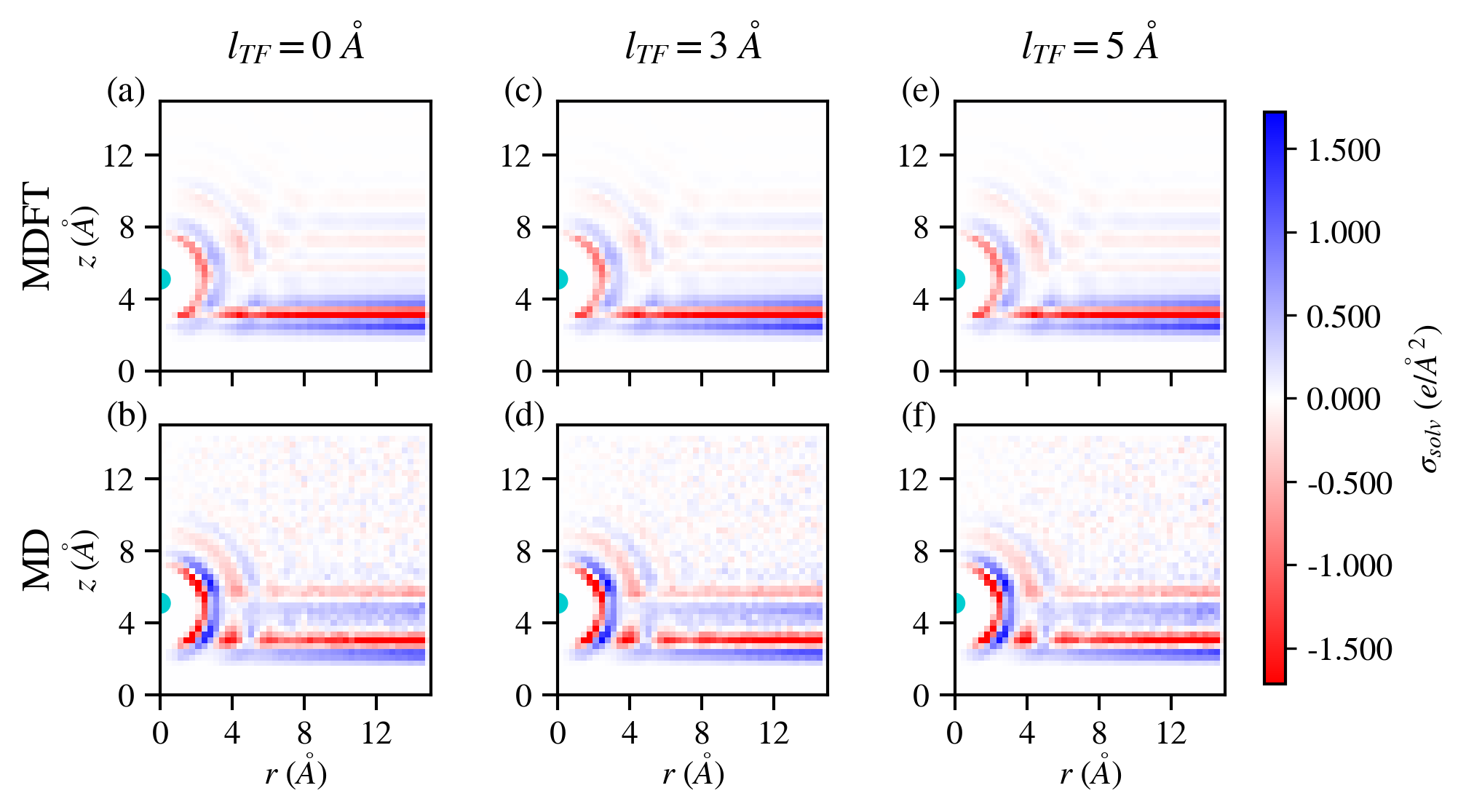}
% with diff: rzmap_Nawat_withdiff.png}% nodiff figure:rzmap_Nawat.png}
\caption{Average charge density of water molecules in the $(r, z)$ plane (see Eq.~\ref{eq:rhosolv}) for the Na$^+$ ion at three different Thomas-Fermi screening lengths ($l_\text{TF}$ = 0, 3, and 5~\AA). Panels (a), (c), and (e) show the MDFT results, while panels (b), (d), and (f) present the MD results.
}
\label{fig:rznawat}
\end{figure*}

Finally, we examine the effect of the Thomas-Fermi screening length on the solvent structure near the ion. To that end, we extend the previous analysis (Fig.~\ref{fig:rzmap}) to three different $l_\text{TF}$ values: 0, 3, and 5~\AA. Fig.~\ref{fig:rznawat} shows the average charge density of water molecules in the $(r,z)$ plane for a Na$^+$ ion at these $l_\text{TF}$ values. We observe no significant change in the solvent charge density when varying $l_\text{TF}$ in both MDFT and MD. This is consistent with the pure solvent case discussed in Section~\ref{sec:results:solvstr:pure}, as well as with the results of Loche \emph{et al.}~\cite{loche_effects_2022} showing a very limited effect of the metallicity of graphite on the potential of mean force for the adsorption of an ion in water. These observations on the solvent structure also hold in acetonitrile, as shown in Fig.~\ref{fig:rznamn}) and or the Cl$^-$ anion in both solvents (see Supplementary Figs.~S2 and S3). In all cases, MDFT consistently captures the asymmetric solvation structure and solvent patterns, demonstrating its robustness in reproducing solvation behavior accurately.

\begin{figure*}[!ht]
\center
\includegraphics[width=6in]{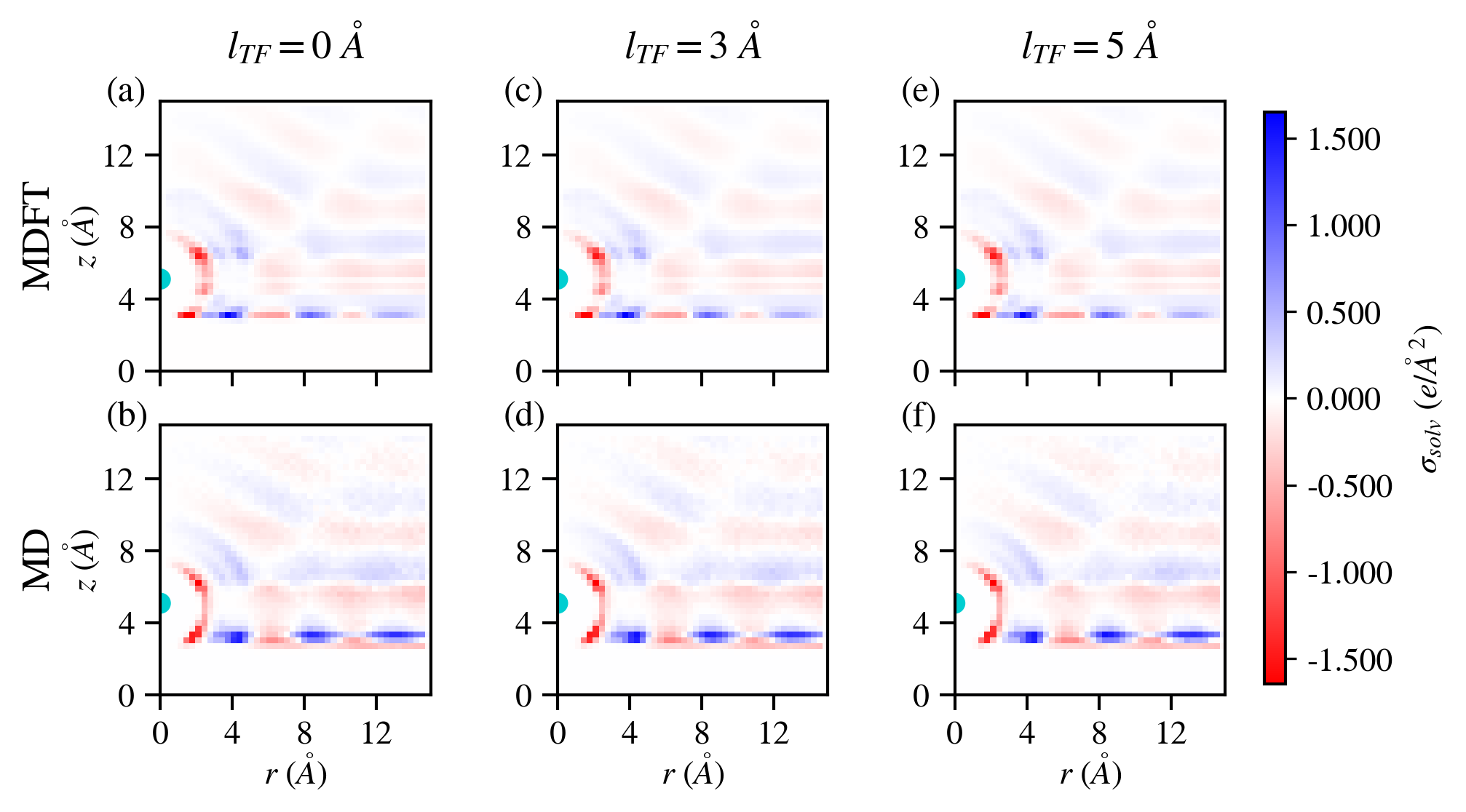}
\caption{Average charge density of acetonitrile molecules in the $(r, z)$ plane (see Eq.~\ref{eq:rhosolv}) for the Na$^+$ ion at three different Thomas-Fermi screening lengths ($l_\text{TF}$ = 0, 3, and 5~\AA). Panels (a), (c), and (e) show the MDFT results, while panels (b), (d), and (f) present the MD results.}
\label{fig:rznamn}
\end{figure*}

Overall, the results from Section~\ref{sec:results:solvstr} indicate that there is no significant effect of $l_\text{TF}$ on the interfacial structure (at least in the absence of applied voltage). This suggests that the effect of $l_\text{TF}$ on the charge distribution induced inside the metal by the ion (Section~\ref{sec:results:inducedq}) is essentially due to how the metal responds to the (same) external charge distribution, even though the coupling between both sides of the interface may play a secondary role.

%%%%%%%%%%%%%%%%%%%%%%%   %%%%%%%%%%%%%%%%%%%%%%%
\section{Conclusion and perspectives}

We investigated the charge induced in a metallic electrode by an ion in solution. Using both classical molecular simulations and molecular Density Functional Theory, we explored the impact of several factors. The effect of the screening inside the metal was studied by tuning the Thomas-Fermi length. The role of the ion, as well as the impact of the solvent, were assessed by considering both a Na$^+$ cation and a Cl$^-$ anion  dissolved in either water or acetonitrile. Consistently with previous molecular-scale studies on the interface between a Thomas-Fermi metal and vacuum or water, we find that at $l_\text{TF}$ = 0~\AA, the induced surface charge is highly localized below the ion and spreads as $l_\text{TF}$ increases.

The main conclusions of the present work are twofold. Firstly, there is no significant effect of the Thomas-Fermi screening length on the interfacial structure (at least in the absence of applied voltage). This suggests that the effect of $l_\text{TF}$ on the charge distribution induced inside the metal by the ion is essentially due to how the metal responds to the  external charge distribution. The coupling between both sides of the interface may play a secondary role. Secondly, MDFT is able to capture accurately the solvation of ions in water and acetonitrile at an atomistically resolved electrode. It reproduces subtle features of the solvent structure at a fraction of the computational cost required to converge the 3D structure with MD simulations. MDFT typically allows for a reduction of the computational cost by three orders of magnitude. As an example, in the pure water case it took roughly 7400~CPUh to obtain 8~ns of simulation, while the functional optimization requested 6~CPUh. While we have focused here on the induced charge and the solvent structure around an interfacial ion, we are currently also exploring the ability of MDFT to reproduce the effect of metallicity on interfacial thermodynamics, which can be investigated with MD as described in Ref~\citenum{scalfi-pnas-2021}.

As discussed in Ref.~\citenum{Nair-2024} for the charge induced in an atomistically resolved electrode by an ion in vacuum, the predictions of  continuum electrostatics capture most of the features observed with the atomistic description (except the oscillations due to the atomic sites of the graphite lattice), provided that the ion-surface distance and the Thomas-Fermi screening length are larger than the inter-atomic distances within the electrode. This requires however a careful definition of the position of the effective interface between the metal and vacuum. The present work provides a molecular basis (both with MD and MDFT) for a systematic quantitative comparison with analytical predictions from continuum electrostatics in the presence of an implicit solvent, either described only by its permittivity~\cite{vorotyntsev_1980} or by more elaborate models interpolating between the long- and short-range screening properties of the solvent~\cite{berthoumieux_dielectric_2019,vatin_electrostatic_2021,monet_nonlocal_2021,hedley_dramatic_2023,hedley_electric_2024}.

Since the induced charges and solvent structure are well reproduced, the next step is to consider thermodynamic and kinetic properties. The potential of mean force as a function of the distance from the surface, for which semi-analytical models have been proposed~\cite{robert_coupled_2023}, is particularly suited for MDFT since the solvation free energy is a direct output of the functional minimization. Moreover we  established a framework to compute Marcus free energy curves with MDFT, allowing to access the activation free energy. MDFT provides a a computationally efficient method to access these quantities, while maintaining a molecular description of the solvent~\cite{Jeanmairet-rsc-2019,Hsu-Marcus-2022}. The present work shows that it should be possible to also include the impact of metallicity by tuning of the Thomas-Fermi length into such such studies, at a more affordable cost than MD simulations~\cite{coello2024microscopic}. This would open the way to a systematic study of the influence of the nature of the ion (size and charge), of the solvent and of the surface (atomic structure and screening length) on adsorption ion and electron transfer kinetics.

%%%%%%%%%%%%%%%%%%%%%%%   %%%%%%%%%%%%%%%%%%%%%%%
\section*{Acknowledgments}
We thank Luc Belloni for useful discussions and for providing us with the MDFT functional for acetonitrile. This project received funding from the European Research Council under the European Union’s Horizon 2020 research and innovation program (grant agreement no. 863473). GJ acknowledges financial support from the French Agence Nationale de la Recherche (project ANR BRIDGE AAP CE29). This work was performed using HPC resources from GENCI-IDRIS (Grant 2024-AD010912966R2). 

%%%%%%%%%%%%%%%%%%%%%%%   %%%%%%%%%%%%%%%%%%%%%%%
\section*{Author declarations}

\subsection*{Conflict of interest}
There is no conflict of interest to declare

\subsection*{Author contributions}

\textbf{Swetha Nair:} Conceptualization (supporting); Formal analysis (equal); Investigation (lead); Methodology (supporting); Software (supporting); Writing/Original Draft Preparation (lead); Writing – review \& editing (equal).
\textbf{Guillaume Jeanmairet:} Conceptualization (equal); Formal analysis (equal); Investigation (supporting); Methodology (equal); Software (lead); Supervision (supporting); Writing – review \& editing (equal).
\textbf{Benjamin Rotenberg:} Conceptualization (equal); Formal analysis (equal); Funding Acquisition (lead); Investigation (supporting); Methodology (equal); Supervision (lead); Writing – review \& editing (equal).

%%%%%%%%%%%%%%%%%%%%%%%   %%%%%%%%%%%%%%%%%%%%%%%
\section*{Data availability}

\todo{To be completed consistently with the final version of the manuscript:} The original data presented in this study are openly available in Zenodo at \todo{[DOI/URL]} or \todo{[reference/accession number]}.

%%%%%%%%%%%%%%%%%%%%%%%  

\newpage

\bibliographystyle{apsrev4-1}
%\bibliography{aipsamp}

%\begin{thebibliography}{10}
%\end{thebibliography}
\providecommand{\noopsort}[1]{}\providecommand{\singleletter}[1]{#1}%
\end{document}